\documentclass[aps,reprint,twocolumn]{revtex4-1}
\usepackage{amsmath} 
\usepackage{amssymb} 
\usepackage{amsfonts}
\usepackage{dsfont}
\usepackage{graphicx} 
\usepackage{color}
\usepackage{subfigure}
\usepackage[latin2]{inputenc}
\usepackage{multirow}
\usepackage[figuresright]{rotating}
\usepackage{hyperref}
\usepackage{units}

\usepackage{floatpag}
\rotfloatpagestyle{empty}

\usepackage[normalem]{ulem}

\begin{document}
\markboth{M\"unnix et. al.}
{Statistical causes for the Epps effect in microstructure noise}

\title{Statistical causes for the Epps effect in microstructure noise}

\author{Michael C. M\"unnix}
\email{michael@muennix.com}
\author{Rudi Sch\"afer}
\author{Thomas Guhr}
\address{Department of Physics, University of Duisburg-Essen 47048 Dusiburg, Germany}

\begin{abstract}
We present two statistical causes for the distortion of correlations on high-frequency financial data. We demonstrate that the asynchrony of trades as well as the decimalization of stock prices has a large impact on the decline of the correlation coefficients towards smaller return intervals (Epps effect). These distortions depend on the properties of the time series and are of purely statistical origin. We are able to present parameter-free compensation methods, which we validate in a model setup. Furthermore, the compensation methods are applied to high-frequency empirical data from the NYSE's TAQ database. A major fraction of the Epps effect can be compensated. The contribution of the presented causes is particularly high for stocks that are traded at low prices.
\end{abstract}


\maketitle
\section{Introduction}
The decline of calculated correlations in financial data towards smaller return intervals was first discovered by Thomas Epps in 1979 \cite{epps79}. This behavior was subsequently detected on different stock exchanges \cite{Bonanno01, kwapien04,zebedee09} and foreign exchange markets \cite{lundin98,muthuswamy01}.
The Epps effect has received considerable attention, from economists as well as from mathematicians and theoretical physicists. 

Hayashi and Yoshida \cite{hayashi05} introduced a cumulative estimator that only considers returns with overlapping time intervals. Hence, it deals with the asynchrony of time series as a cause for the Epps effect. Subsequently, Voev and Lunde \cite{voev07} demonstrated that this estimator can be biased in the presence of noise and proposed a bias correction. Griffin and Oomen \cite{griffin06} extended the estimator of Hayashi and Yoshida by adjustments for lagged correlations. The work of T\'oth and Kert\'esz \cite{toth09} also deals with the phenomenon of lagged correlations. They introduce a model that is based on the decomposition of cross-correlations.
The recent study of Zhang \cite{zhang08}, shows that usual previous-tick-estimators are biased. They consequently provide an optimal sampling frequency of returns to suppress the Epps effect. Barndorff-Nielson et. al. \cite{barndorff02,barndorff08} examine high frequency correlations and propose multivariate realized kernels to significantly improve the estimation of correlations.
An extensive study of microscopic causes leading to the Epps effect has been performed by Ren\`o \cite{reno03}.

Clearly, many mechanisms contribute to the Epps effect. We demonstrate that there are two major causes of purely statistical origin. Our aim is not to develop a complete description of the Epps effect. We rather want to identify statistical causes that can be compensated directly, without the requirement of adjusting parameters,  model calibrations or an optimal sampling frequency.  The two major causes we identify are the asynchrony of the time series and the impact of the decimalization by the tick-size.

This paper is organized as follows. In section \ref{s:rev-async} and \ref{s:rev-tick}, we present compensation methods for the impact of asynchronous time series and the impact of the tick-size. This is followed by a combined compensation of both effects in section \ref{s:rev-combcomp}.
The results are validated in a model setup in section \ref{s:rev-model}. In section \ref{s:rev-empiric}, we apply the compensation methods to empirical data from the NYSE's TAQ database to estimate the impact of the statistical causes on the Epps effect. We discuss our results in section \ref{s:rev-conc}.

\section{Theory of compensating microstructure noise distortions}
In the sequel, we give an overview over compensation methods that account for distortions of the Pearson correlation coefficient due to statistical effects. In particular, the asynchrony of trading times and the impact of the tick-size are considered.
\subsection{Asynchrony of trading times}
\label{s:rev-async}
We begin with demonstrating how the asynchrony of time series contributes to the Epps effect. By asynchrony we refer to time series that feature an arbitrary lag for a given point in time but the average lag is zero. The asynchrony is simply due to the non-synchronous pricing of stocks. A detailed derivation and study of our finding is performed in \cite{muennix09b}

T\'oth and Kert\'esz \cite{toth09} stated that the impact of the asynchrony is weak, compared to the impact of a static lag, for which they developed a model. In the following, we will demonstrate that the asynchrony can be a major cause for the Epps effect. 

\begin{figure}[tb]
\begin{center}
\includegraphics[width=1\columnwidth]{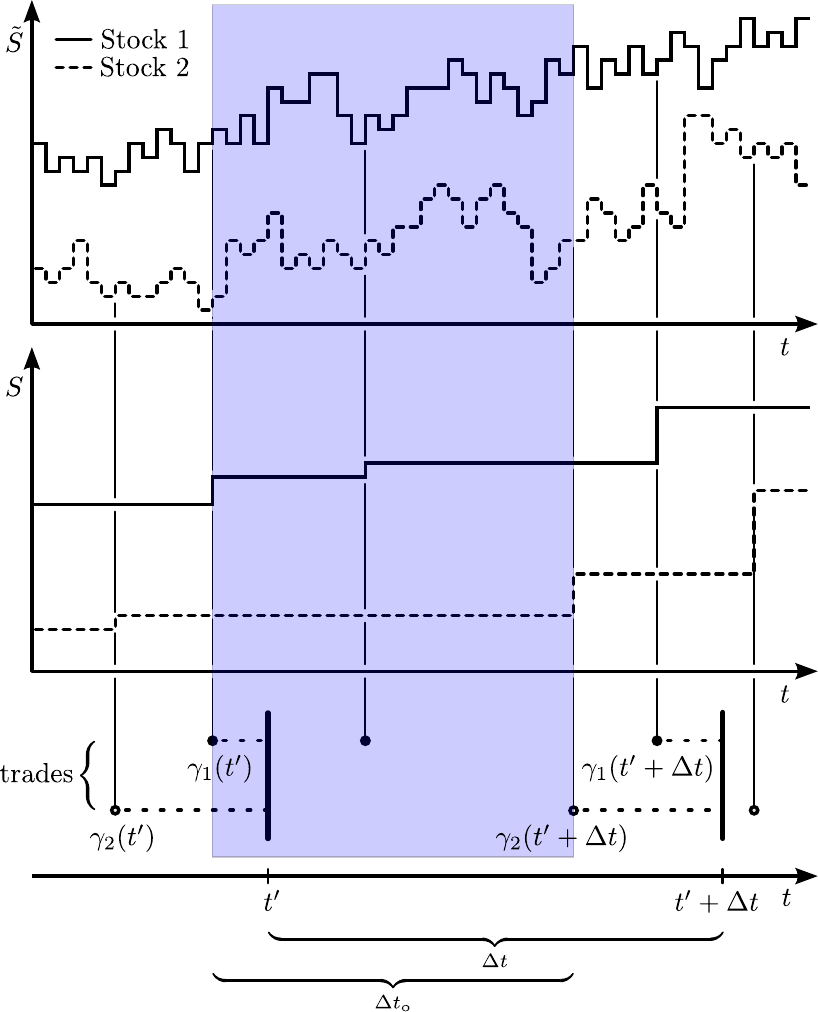}
\caption{Illustration of the model for asynchronous trading times of two stocks. Shown on the top are the prices $\tilde{S}$ for the hypothetical underlying timescale. The ``sampling'' of theses prices to macroscopic prices $S$ with randomly distributed points of trades is illustrated in the middle. The points of trades are indicated by the vertical lines. The thick bars at the bottom illustrate the return interval between $t'$ and $t'+\Delta t$. The points of last trades on these times are denoted with $\gamma$. The shaded area indicates the overlap $\Delta t_{\mathrm{o}}$.}
\label{fig:rev-overlap}
\end{center}
\end{figure}

The central assumption of this approach is the existence of underlying non-lagged time series of prices. The assumption of a finer (see, e.g., \cite{toth09}) or continuous (see, e.g., (\cite{hayashi05,barndorff02,reno03}) timescale is frequently used in the estimation of correlations. This ansatz is also intuitive, as most stocks are traded at several stock exchanges and off-exchange (OTC) simultaneously. 
The basic idea is the following: Due to the asynchrony, each term of the Pearson correlation coefficient can be divided into a part which contributes to the correlation and a part which is uncorrelated and therefore lowers the correlation coefficient.

\begin{figure}[b]
\begin{center}
\includegraphics[width=1\columnwidth]{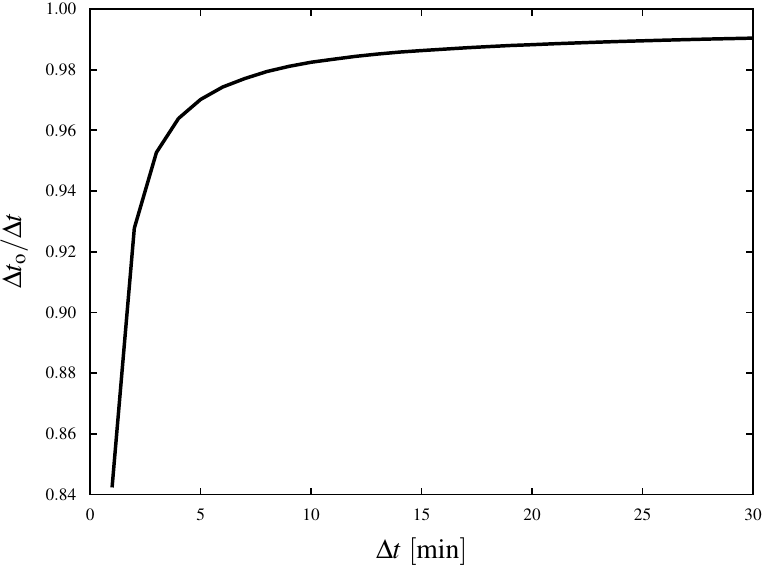}
\caption{Average fractional overlap for the 5 highest correlated stock pairs of each industry branch in the S\&P 500 index versus the return interval $\Delta t$.}
\label{fig:rev-fractionaloverlap}
\end{center}
\end{figure}

The situation is sketched in Fig. \ref{fig:rev-overlap}. Here, $\gamma_{i}(t)$ refers to the point of last trade, for the $i$-th stock at time $t$. The waiting times, that is, the periods between two consecutive trades, are randomly (usually exponentially) distributed. Hence, when calculating the return of the interval from $t'$ to $t'+\Delta t$, we actually obtain the return on an effective return interval which is between the points of last trade referring to the right and the left side of the return interval, that are $[\gamma_{1}(t');\gamma_{1}(t'+\Delta t]$ and $[\gamma_{2}(t');\gamma_{2}(t'+\Delta t]$. These intervals can be smaller or larger than the initially chosen return interval. When considering the returns of two stocks within the same interval, one obtains two effective return intervals that are in most cases not equal in length, start-point and end-point. These intervals usually share an overlap $\Delta{t}_{\mathrm{o}}$, although this is not necessarily true for stocks that are traded on low quantities. This overlap is given by
\begin{eqnarray}
\Delta{t}_{\mathrm{o}}(t')&=&\min\left(\gamma_{1}(t'+\Delta t),\gamma_{2}(t'+\Delta t)\right)\nonumber\\
&&-\max\left(\gamma_{1}(t'),\gamma_{2}(t')\right)\ .
\end{eqnarray} 
The fractional overlap $\Delta{t}_{\mathrm{o}}/\Delta t$ declines with lower return intervals as shown in Fig. \ref{fig:rev-fractionaloverlap}. This scaling behavior already looks similar to the Epps effect on correlation coefficients. As we will demonstrate, the fractional overlap is strongly connected to the Epps effect.

In the sequel, we consider the Epps effect on relative price changes or arithmetic financial returns $r$ which are defined as the relative price change during a return interval $\Delta t$. It reads

\begin{equation}
r(t) = \frac{S(t+\Delta t)-S(t)}{S(t)} \ ,
\label{eq:rev-return}
\end{equation} 
where $S(t)$ refers to the price of a security at time $t$.

Regarding the hypothetical underlying time series, the information within the overlap $\Delta{t}_{\mathrm{o}}$ is synchronous. This part gives the \emph{true} correlation of the time series.
In contrast, the parts left and right from the overlap are asynchronous. Under the assumption of randomly distributed trading times, these parts are on average uncorrelated. Hence, the returns outside the overlap distort the correlation coefficient.

It follows that the contribution of these two single returns to the Pearson correlation coefficient is the Pearson correlation coefficient of the underlying time series multiplied by the fractional overlap as shown in \cite{muennix09b}.
This can easily be outlined, when considering the normalized returns of the underlying time series
\begin{eqnarray}
\tilde g(t) = \frac{\tilde r(t)-\left\langle \tilde r \right\rangle}{\sigma_{\tilde r}} \ .
\end{eqnarray}
Here, $\langle \cdots \rangle$ denotes the average over $T$ and $\sigma$ denotes to the standard deviation of the time series with length $T$. The tilde indicates the underlying time series. When calculating the Pearson correlation coefficient of two time series, the overlap is a function of the time step: $\Delta{t}_{\mathrm{o}}=\Delta{t}_{\mathrm{o}}(t)$. We denote the interval of the overlap $\Delta t_{\mathrm{o}}(t)$ for each time step as $\mathcal J(t)$. The time steps on the underlying timescale are denoted with $\tilde t$. We can re-arrange the terms of the correlation coefficient in terms that originate from within the overlap-interval and thus are synchronous, and terms that are asynchronous, 
\begin{eqnarray}
\mathrm{corr}(r_{1},r_{2})\propto
\frac{1}{T}\sum\limits_{t=0}^{T}
\left(\left( \underbrace{\sum\limits_{\tilde t \notin \mathcal J(t)}{\tilde{g}_{1}(\tilde t)}}_{\mathrm{async.}} +  \underbrace{\sum\limits_{\tilde t \in \mathcal J(t)}{\tilde{g}_{1}(\tilde t)}}_{\mathrm{sync.}} \right)\right. \nonumber\\
\left.\times\left( \underbrace{\sum\limits_{\tilde t \notin \mathcal J(t)}{\tilde{g}_{2}(\tilde t)}}_{\mathrm{async.}} +  \underbrace{\sum\limits_{\tilde t \in \mathcal J(t)}{\tilde{g}_{2}(\tilde t)}}_{\mathrm{sync.}}\right) \right) \ .
\end{eqnarray}
This leads to
\begin{eqnarray}
\mathrm{corr}(r_{1},r_{2})
&=& \frac{1}{T}\sum\limits_{t=0}^{T}\mathrm{corr}_{t}(\tilde{g}_{1},\tilde{g}_{2}) \frac{\Delta{t}_{\mathrm{o}}(t)}{\Delta{t}}\ ,
\label{rev-result}
\end{eqnarray}
where, $\mathrm{corr}_{t}(\tilde{g}_{1},\tilde{g}_{2})$ is the Pearson correlation coefficient of the underlying time series for the interval $[t,t+\Delta t]$. It gives the \emph{true} correlation. Each term of the correlation coefficient is multiplied by the fractional overlap $\Delta t / \Delta{t}_{\mathrm{o}}(t)$, because only the information inside the overlap contributes to the correlation coefficient. 

As we are able to quantify the impact of the fractional overlap on the correlation coefficient, we can easily compensate this distortion by
\begin{equation}
\mathrm{\widehat{corr}}_{\mathrm{async}}(r_{1},r_{2})  = \left\langle{g}_{1}(t_{j}){g}_{2}(t_{j}) \frac{\Delta{t}}{\Delta{t}_{\mathrm{o}}(t_{j})} \right\rangle\ ,
\label{eq:rev-corr-cor}
\end{equation}
where $g$ refers to the normalized return of the corresponding (non-hypothetical) time series.
Furthermore, only returns should be considered that actually share an overlap, analogously to the estimator of Hayashi and Yoshida \cite{hayashi05}. 

Initially, we made the assumption of an underlying time series of prices, which is correlated and which exists on a smaller time scale.
Equation (\ref{eq:rev-corr-cor}) does no longer depend on the time scale of the hypothetical underlying time series. Neither does it depend on the actual prices on the underlying time series. Hence, the only necessary assumption is that there exists underlying \emph{information} which is correlated on a finer time scale. This is an important finding, since Martens and Poon \cite{martens01} indicated that the synchronization of returns from international stock exchanges is a non-trivial problem.

\subsection{Tick-Size}
\label{s:rev-tick}
We now turn to the second statistical cause. We estimate the tick-size's impact on the Epps effect. A comprehensive derivation and discussion is performed in \cite{muennix10a}.

The lowest possible price change, the \emph{tick-size}, of most securities has been constantly reduced, resulting in tick-sizes of \nicefrac{1}{100}-th of the respective currency on most stock exchanges. This process is often referred to as decimalization. It was, e.g., motivated by aiming at an enhanced market efficiency. In theory, small tick-sizes allow for a faster clearing of market arbitrage. However, the question whether a smaller tick-size generally improves the market quality is controversially discussed \cite{declerck02,huang01}.
Among others, Harris \cite{harris91} discussed that a larger tick-size can ensure liquidity, but on the other hand, it can lead to erroneous data in financial indices \cite{kozicki09}.
In this context, Angel \cite{angel97} observed that the prices of a stock are commonly in a typical range, which is optimal to provide liquidity. Companies perform stock splits to control the absolute price of their share.
A recent study by Onnela et. al. \cite{onnela09} indicates that in some cases only a fraction of the theoretically possible prices are used. Hence, prices cluster at certain multiples of the tick-size resulting in an effective tick-size.

\begin{figure}[tb]
\begin{center}
\includegraphics[width=1\columnwidth]{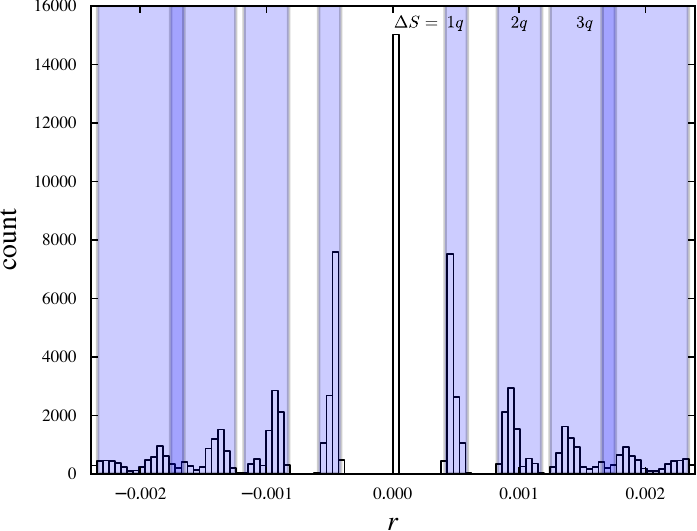}
\caption{Detail of the distribution of 5-min returns of the AES Corp. share in 2007. The shaded areas refer to returns that originate from the same absolute price change $\Delta S$. The price changes $\Delta S$ are denoted as multiplies of the tick-size $q$.}
\label{fig:rev-returns}
\end{center}
\end{figure}

At first glance, it could appear that the transition from absolute price changes $\Delta S$ to returns $r$ removes this discretization from the distribution, since the returns are almost continuously distributed. A detailed look at the center of a return distribution (see Fig. \ref{fig:rev-returns}) reveals that the discretization effects are still visible. Despite its poor graphical visibility, the discretization affects returns on all intervals. Especially, we expect an impact on the correlation coefficient if the discretization is high, that is, when stocks are traded at low prices.

Due to the imposing of discrete prices, information is lost. The average relative price change becomes smaller when considering smaller return intervals. The tick-size remains the same of course. The information loss grows for smaller return intervals. Thus, the discretization should also contribute to the Epps effect.

The basic assumption of our model is that we can statistically describe the discreteness in market prices by a discretization of a hypothetical underlying price. 
Of course, the market prices do not actually result from a discretization process. However, there is a large variety of trading strategies simultaneously acting on the market. These strategies also act on a large spectrum of different investment horizons. There are even traders that try to exploit the finite tick-site in their trading strategies. 
As the price formation results from the interaction of this diversity of trading strategies, the price fluctuations on the level of the tick-size can be viewed as purely statistical. Hence, a natural approach is the assumption of on average uniformly distributed discretization errors.

Using the arithmetic return defined in equation (\ref{eq:rev-return}), we introduce the discretization error $\vartheta$ as
\begin{eqnarray}
\bar r(t)&=&\frac{\Delta \bar S(t)}{\bar S(t)}\\
r(t)&=&\frac{\Delta \bar S(t)+ \vartheta}{\bar S(t)}
\label{eq:rev-disc}
\end{eqnarray}
with
\begin{eqnarray}
\Delta \bar S(t) = \bar S(t) - \bar S(t+\Delta t) \ ,
\end{eqnarray}
where $\bar S(t)$ denotes to the discretized stock price and $\bar r(t)$ denotes the return, which is based on discretized stock prices. We emphasize that we do not account for the discretization of the prices $S(t)$, but consider only the discretization of the price changes $\Delta S(t)$. We demonstrate in section \ref{s:rev-model} that this only induces a negligible error. As $\vartheta$ in equation (\ref{eq:rev-disc}) is actually the difference of two uniformly distributed discretization errors, it follows a triangular distribution.

The calculation of the Pearson correlation coefficient including the discretization errors as introduced in (\ref{eq:rev-disc}) leads to
\begin{eqnarray}
\mathrm{\widehat{corr}}_{\mathrm{tick}}(r_1,r_2)&=&
\frac{\mathrm{cov}(r_1,r_2)}{\sigma_{r_{1}}\sigma_{r_{2}}}\\
&=&
\bigg[
\mathrm{cov}\left(\bar{r}_1,\bar{r}_2\right)
\nonumber\\
&&
\mathrm{cov}\left(\frac{\Delta \bar{S}_1}{S_1},\frac{\vartheta_{2}}{S_2}\right)+
\mathrm{cov}\left(\frac{\Delta \bar{S}_2}{S_2},\frac{\vartheta_{1}}{S_1}\right)
\nonumber\\
&&+
\mathrm{cov}\left(\frac{\vartheta_{1}}{S_1},\frac{\vartheta_{2}}{S_2}\right) 
\bigg]\bigg/(\hat \sigma_{r_{1}}\hat \sigma_{r_{2}})
\label{eq:rev-tickcomp}
\end{eqnarray}
with

\begin{eqnarray}
\hat \sigma_{r_{i}}=
\sqrt{\mathrm{var}\left(\bar{r}_i\right)+\mathrm{var}\left(\frac{\vartheta_{i}}{S_i}\right)+2\mathrm{cov}\left(\frac{\Delta \bar{S}_i}{S_i},\frac{\vartheta_{i}}{S_i}\right)} \ .
\label{eq:rev-discsigma}
\end{eqnarray}
Here, $\bar r$ refers to the return with respect to discretized prices. Only the terms $\mathrm{cov}\left(\bar{r}_1,\bar{r}_2\right)$, $\mathrm{var}\left(\bar{r}_1\right)$ and $\mathrm{var}\left(\bar{r}_2\right)$ can be calculated with the discretized prices. All other terms are unknown and describe the information loss due to the discretization. We can estimate these terms and thereby compensate for the information loss by interpolating the price change distribution. Technically, this is achieved by expanding the variance and covariance terms in equation (\ref{eq:rev-tickcomp}) and (\ref{eq:rev-discsigma}) and estimating the discretization error for all price changes individually. Estimation techniques for the individual discretization errors are comprehensively discussed in \cite{muennix10a}.
This study indicates that only certain terms of equation (\ref{eq:rev-tickcomp}) have a noticeable impact on the compensation. If calculation speed is an issue, one can approximate
\begin{eqnarray}
\mathrm{\widehat{corr}}_{\mathrm{tick}}(r_1,r_2) \approx
\frac{\mathrm{cov}\left(\bar{r}_1,\bar{r}_2\right)}
{\hat \sigma_{r_{1}}\hat \sigma_{r_{2}}}\ .
\label{eq:rev-tickcompapprox}
\end{eqnarray}
The main contribution to the distortion of correlation coefficients in small return intervals is the overestimation of $\sigma$. Fig. \ref{fig:rev-realized-vola} in section \ref{s:rev-model} shows this overestimated $\sigma$ and the tick-size-corrected $\hat \sigma$ versus the return interval $\Delta t$.
This is consistent with the findings of Hansen and Lunde \cite{hansen06}. They demonstrate that the realized variance is overestimated on small return intervals due to microstructure noise. The empirical evidence in section \ref{s:rev-empiric} indicates that the tick-size have profound impact on this noise.

Due to the convex shape of the price change distribution, the discretization errors are not distributed symmetrically. This effect grows with the impact of the discretization, i.e., smaller return intervals. Thus, the estimation of variances on the discretized values is biased. This gives the largest contribution to the distortion of correlation coefficients due to discretized data. We can correct this behavior with the presented compensation.
\subsection{Combined compensation}
\label{s:rev-combcomp}
\begin{figure}[tb]
\begin{center}
\includegraphics[width=1\columnwidth]{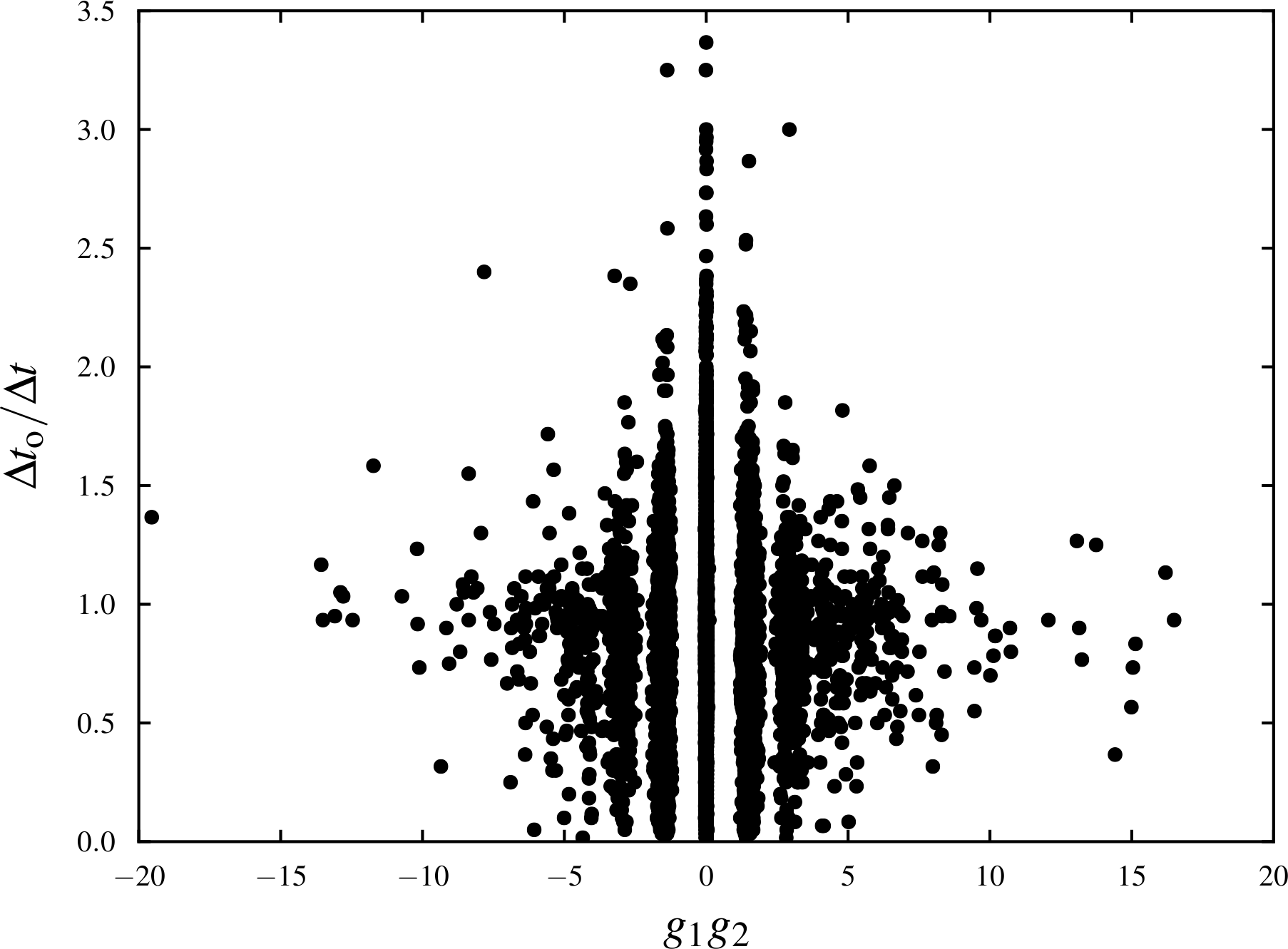}
\caption{Product of normalized return pairs versus fractional their fractional overlap for 1-min returns of the shares of Novell Inc. and Unisys Corp.in 2007. The average fractional overlap is 0.76.}
\label{fig:rev-ovsr}
\end{center}
\end{figure}

Having presented compensation methods for distortions of the correlation coefficient due to asynchronous time series and due to the tick-size, we now combine both findings. The compensation of asynchrony acts on each term of the Pearson correlation coefficient for every point in time. The tick-size compensation, in contrast, acts on the Pearson correlation coefficient as a whole in terms of the time series, but it acts on every occurring price change individually. Both effects superimpose, as illustrated in Fig. \ref{fig:rev-ovsr}. The horizontal axis shows the product of normalized 1-min returns for each point in 2007 (overnight returns are excluded). The vertical axis shows the corresponding fractional overlap of each return pair. The discretization effects are visible in the center, superimposed with the asynchronous characteristics. Similar to the findings of Szpiro \cite{szpiro98} for single stocks, the tick-size induces a nanostructure on the terms of the Person correlation coefficient for two stocks. 

The simultaneous compensation of both effects can be achieved by combining both presented compensations. It reads
\begin{eqnarray}
\mathrm{\widehat{corr}}(r_1,r_2)
&=&
\left\langle  \bar{r}_1 \bar{r}_2\frac{\Delta t}{\Delta t_{\mathrm{o}}} \right\rangle
\\&&+
\bigg(
\mathrm{cov}\left(\frac{\Delta \bar{S}_1}{S_1},\frac{\vartheta_{2}}{S_2}\right)+\mathrm{cov}\left(\frac{\Delta \bar{S}_2}{S_2},\frac{\vartheta_{1}}{S_1}\right)
\nonumber\\&&+
\mathrm{cov}\left(\frac{\vartheta_{1}}{S_1},\frac{\vartheta_{2}}{S_2}\right)
- \left\langle \bar{r}_1 \right\rangle \left\langle \bar{r}_2 \right\rangle
\bigg)
\nonumber\\&&\times
\left\langle \frac{\Delta t}{\Delta t_{\mathrm{o}}} \right\rangle 
\bigg / (\hat \sigma_{r_{1}} \hat \sigma_{r_{2}})
\ .
\end{eqnarray}
Analogously to the previous section, this expression can be approximated by 
\begin{eqnarray}
\mathrm{\widehat{corr}}(r_1,r_2)
&\approx&
\frac{
\left\langle  \bar{r}_1 \bar{r}_2\frac{\Delta t}{\Delta t_{\mathrm{o}}} \right\rangle  
}
{\hat \sigma_{r_{1}}\hat \sigma_{r_{2}}}\ .
\end{eqnarray}
By multiplying the covariance terms of discretized returns $\bar r$ with the inverse fractional ${\Delta t}/{\Delta t_{\mathrm{o}}}$ overlap and by correcting the overestimation of the standard deviations $\sigma$, the largest fraction of the correlation coefficient's distortion can be compensated.

\section{Results}
Before applying the method to empirical data, we study it in a model setup. Subsequently, we apply the compensation methods to empirical data from the NYSE's TAQ database to estimate the impact of the presented causes on the distortion of correlation coefficients.
\subsection{Model results}
\label{s:rev-model}
We start by generating an underlying correlated time series using a GARCH(1,1) model, as introduced in \cite{bollerslev86}
\begin{equation}
r_{i}(t) = \sigma_{i}(t)\left( \sqrt{c}\,\eta(t) + \sqrt{1-c}\,\varepsilon_{i}(t)\right)\ .
\end{equation}
Here $r_{i}(t)$ stands for the return of the $i$-th stock at time $t$ and $c$ is the correlation coefficient. The random variables $\eta(t)$ and $\varepsilon_{i}(t)$ are taken from standard normal distributions. $\eta(t)$ is identical for all stocks; It induces the correlation. The $\varepsilon_{i}$ are individual for each stock.
$\sigma_i(t)$ is the non-constant variance, given by a GARCH(1,1) process
\begin{equation}
\sigma_{i}^{2}(t)=\alpha_{0}+\alpha_{1}r_{i}^{2}(t-1)+\beta_{1}\sigma_{i}^{2}(t-1) \ .
\end{equation}
The initial parameters of the GARCH(1,1) process are chosen as $\alpha_{0}=2.4	 \times 10^{-4}$, $\alpha_{1}=0.15$ and $\beta_{1}=0.84$.

\begin{figure}[tb]
\begin{center}
\includegraphics[width=1\columnwidth]{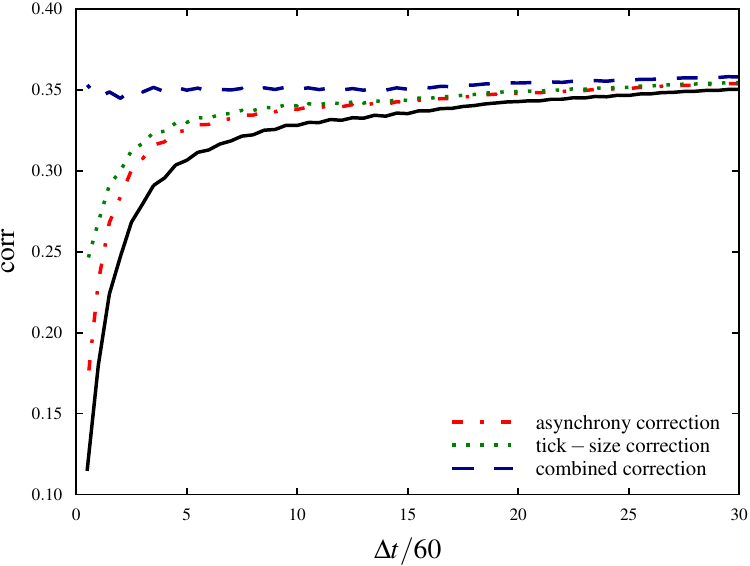}
\caption{Model results of compensation methods.}
\label{fig:rev-modelresults}
\end{center}
\end{figure}

Two return time series $r_{1}$ and $r_{2}$ are generated representing two correlated stocks. The total lengths of these time series is chosen as $7.2\times10^{6}$, corresponding to a return interval $\Delta t$ of one second during one trading year.
From these returns, we generate two underlying price time series $\tilde S_{1}$ and $\tilde S_{2}$.
We set the starting prices to $t=0$ to $1000$. $c$ is chosen as 0.4.

To model the asynchronous trade processes, these prices are sampled independently using exponentially distributed waiting times with average values typical for the stock market. We choose the average waiting times as 15 and 25 data points (equivalent to seconds in this setup).
In the next step, we round the prices to integer values. An integer price of, for example, 1000 then corresponds to a price of 10 and a tick-size of 0.01. 

Finally, we construct the time series of returns from these prices using return intervals from 60 data points (corresponding to 1 minute) to 1800 data points (corresponding to 30 minutes). The thus obtained time series features both, asynchrony and discretization.

\begin{figure}[tb]
\begin{center}
\includegraphics[width=1\columnwidth]{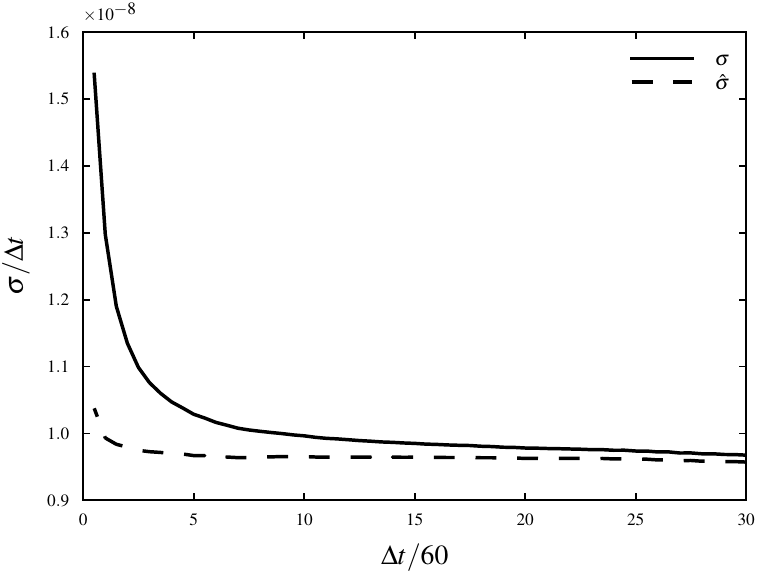}
\caption{Overestimated standard deviation $\sigma$ and tick-size-corrected standard deviation $\hat \sigma$ versus the return interval $\Delta t$ within the model.}
\label{fig:rev-realized-vola}
\end{center}
\end{figure}

The results of the applied compensation methods are shown in Fig \ref{fig:rev-modelresults}.
We are able to compensate the statistical distortion of correlation coefficients almost completely.
The remaining decline of the corrected correlation coefficient on very small return intervals is due to the approximations presented in sections \ref{s:rev-tick} and \ref{s:rev-combcomp} (only price change discretization is considered) as well as the negligence of the correlation between price changes and prices and the discretization of prices. The impact of the overestimation of the standard deviation $\sigma$ is shown in Fig. \ref{fig:rev-realized-vola}. This illustrates that the tick-size can have a large impact on the overestimation of $\sigma$. Moreover, we are able to compensate for this behavior down to approximately $\Delta t = 180$ time steps (corresponding to 3 minutes in our model).

\subsection{Empirical evidence}
\label{s:rev-empiric}

As already mentioned, many mechanisms contribute to the Epps effect. Our present aim is to quantify the part, which is caused by the statistical properties of the time series.

It is difficult to isolate the Epps effect on single stock pairs, as it is superimposed with other effects leading to other characteristics of the correlation coefficient than expected for the Epps effect.
\begin{figure}[tbp]
\begin{center}
\subfigure[Asynchrony compensation]{
\includegraphics[width=1\columnwidth]{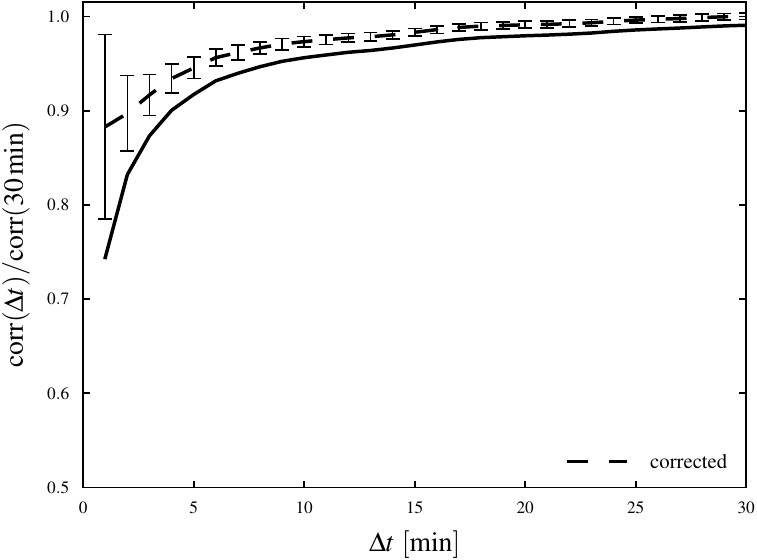}
\label{fig:rev-res-aync}
}
\subfigure[Tick-size compensation]{
\includegraphics[width=1\columnwidth]{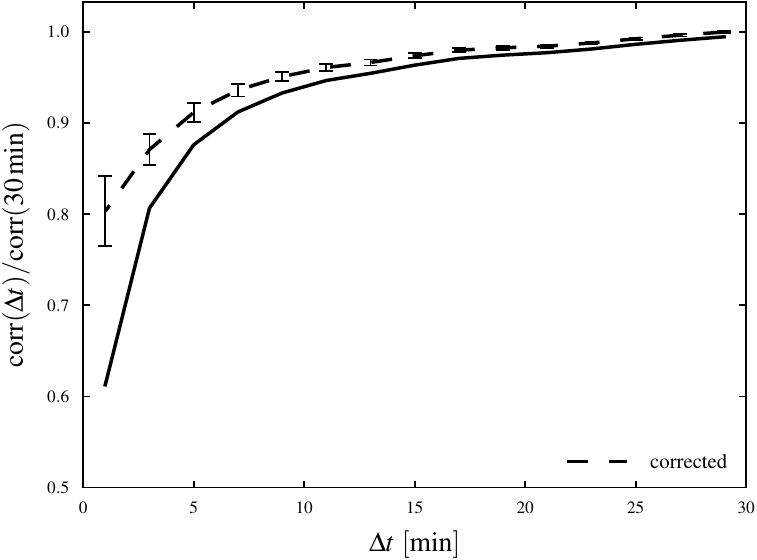}
\label{fig:rev-res-tick}
}
\subfigure[Combined compensation]{
\includegraphics[width=1\columnwidth]{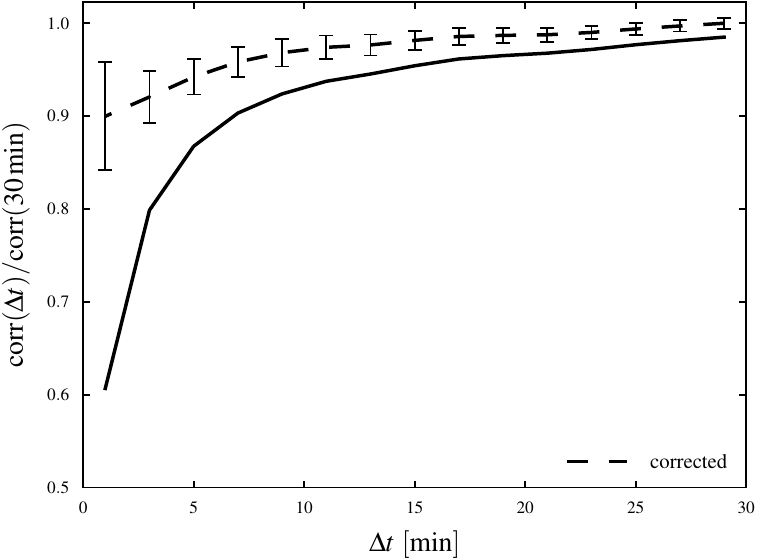}
\label{fig:rev-res-combo}
}
\caption{Empirical results of applied compensation methods. a) represents the average over the 50 highest correlated stock pairs in the S\&P 500 index in 2007 (Top 5 from each industry branch). b) and c) represent the average over the 25 highest correlated stock pairs that were traded between \$10.01 and \$20.00 in 2007. The correlation coefficients have been individually normalized to the corrected value at $\Delta t$ = 30 min}
\label{fig:rev-empi}
\end{center}
\end{figure}

Because of that, we classify two ensembles of stock pairs. After compensating the asynchrony effect for each pair, we build the average for the ensemble by normalizing the correlation coefficients individually by their saturation value at a return interval of 30 minutes.
We also plot the error bars of the compensation representing the double standard deviation $2\sigma$ of the correction. By this method, we can show the scope of the asynchrony model and identify regions, in which other effects dominate. All data is extracted from the NYSE's TAQ database for the year 2007. 

The first ensemble consists of 5 stock pairs of each industry sector of the S\&P 500 index (50 stocks in total), whose daily returns provide the strongest correlation during the year 2007.
We applied the asynchrony compensation to this ensemble.
The results shown in Fig \ref{fig:rev-res-aync} indicate that the asynchrony has a pronounced impact on the Epps effect. It appears that asynchrony effects are the dominating cause for the Epps effect on return intervals down to  approximately 10 minutes, where the remaining Epps effect is on average less than 3\% of the correlation coefficient's saturation value at large return intervals.
Of course, within the statistical ensemble stock pairs can be found which either do not show an Epps effect or which are so infrequently traded that the assumption of an underlying timeline might be unreasonable. Even though the assumption of an underlying time series is a common and intuitive approach, it may not be valid for stocks traded on very low frequencies.

The second ensemble serves as a test scenario for the tick-size compensation. We expect the tick-size to only have a large impact on the correlation coefficient, if the discretization is high, i.e., if stocks are traded at low prices. Thus, we construct the second ensemble out of the 25 most strongly correlated stocks in the S\&P 500 that are traded in the range of \$10.01 to \$20.00.
The price change distributions are segment-wise interpolated with heavy tailed distributions as, i.e., suggested by \cite{b_cont}.
The results in Fig. \ref{fig:rev-res-tick} indicate that for stocks that are traded at low prices, the tick-size can have a sizable impact on the Epps effect.

Eventually, we apply a combined compensation to the second ensemble. Results are shown in Fig. \ref{fig:rev-res-combo}. The empirical evidence indicates that statistical effects can have a very profound impact on the Epps effect.

\section{Concusions}
\label{s:rev-conc}
We demonstrated that statistical causes can have a large impact on the Epps effect, especially for stocks that are traded at low prices. The asynchrony of time series as well as the tick-size have a major impact on the Epps effect. We developed two simple methods to compensate for these causes.

However, this is not a full description of the Epps effect as there are certainly many phenomena contributing to it. In certain scenarios, other statistical properties of the time series or other causes for the Epps effect might dominate. The size of the error bars in Fig. \ref{fig:rev-empi} indicates that the asynchrony compensation does not give reliable results for return intervals below 3 minutes. Especially for very small return intervals, a lag between the time series of two stocks might be the dominating cause, as suggested by T\'oth and Kert\'esz \cite{toth09}.

For stocks that are infrequently traded at very low prices (often referred to as \emph{penny-stocks}) the assumption of uniformly distributed discretization errors needs to be carefully reflected. It is possible that certain trading strategies dominate for those stocks leading to an asymmetrical distribution of discretization errors. 

Nonetheless, the presented compensations significantly improve the estimation of financial correlations. These methods do not require parameter adjustments or model calibrations. Our empirical study indicates that the identified causes can contribute up to 75\% of the Epss effect for stocks that are traded at low prices.

\section*{Acknowledgements}
M.C.M. acknowledges financial support from the Fulbright program and from Studienstiftung des deutschen Volkes.

\bibliographystyle{elsarticle-num}
\bibliography{Manuscript.bbl}

\begin{thebibliography}{10}
\expandafter\ifx\csname url\endcsname\relax
  \def\url#1{\texttt{#1}}\fi
\expandafter\ifx\csname urlprefix\endcsname\relax\def\urlprefix{URL }\fi
\expandafter\ifx\csname href\endcsname\relax
  \def\href#1#2{#2} \def\path#1{#1}\fi

\bibitem{epps79}
T.~W. Epps, \href{http://www.jstor.org/stable/2286325}{Comovements in stock
  prices in the very short run}, Journal of the American Statistical
  Association 74~(366) (1979) 291--298.
\newline\urlprefix\url{http://www.jstor.org/stable/2286325}

\bibitem{Bonanno01}
G.~Bonanno, F.~Lillo, R.~N. Mantegna, High-frequency cross-correlation in a set
  of stocks, Quantitative Finance 1~(1) (2001) 96--104.

\bibitem{kwapien04}
J.~Kwapien, S.~Drozdz, J.~Speth,
  \href{http://www.sciencedirect.com/science/article/B6TVG-4BRJY2K-1/2/d141755%
cb92c5d1b33cead1bb57b0bed}{Time scales involved in emergent market coherence},
  Physica A 337~(1-2) (2004) 231 -- 242.
\newblock \href {http://dx.doi.org/10.1016/j.physa.2004.01.050}
  {\path{doi:10.1016/j.physa.2004.01.050}}.
\newline\urlprefix\url{http://www.sciencedirect.com/science/article/B6TVG-4BRJ%
Y2K-1/2/d141755cb92c5d1b33cead1bb57b0bed}

\bibitem{zebedee09}
A.~A. Zebedee, M.~Kasch-Haroutounian,
  \href{http://www.sciencedirect.com/science/article/B6V7T-4TX794N-1/2/26edc1f%
881adc4aad35511cb584cb8d2}{A closer look at co-movements among stock returns},
  Journal of Economics and Business 61~(4) (2009) 279 -- 294.
\newblock \href {http://dx.doi.org/10.1016/j.jeconbus.2008.11.001}
  {\path{doi:10.1016/j.jeconbus.2008.11.001}}.
\newline\urlprefix\url{http://www.sciencedirect.com/science/article/B6V7T-4TX7%
94N-1/2/26edc1f881adc4aad35511cb584cb8d2}

\bibitem{lundin98}
M.~Lundin, M.~Dacorogna, U.~A. M{\"u}ller, Financial Markets Tick By Tick,
  Wiley, 1998, Ch. Correlation of High Frequency Financial Time Series.

\bibitem{muthuswamy01}
J.~Muthuswamy, S.~Sarkar, A.~Low, E.~Terry, Time variation in the correlation
  structure of exchange rates: high-frequency analyses, Journal of Futures
  Markets 21~(2) (2001) 127--144.

\bibitem{hayashi05}
T.~Hayashi, N.~Yoshida, On covariance estimation of non-synchronously observed
  diffusion processes, Bernoulli 11~(2) (2005) 359--379.

\bibitem{voev07}
V.~Voev, A.~Lunde,
  \href{http://jfec.oxfordjournals.org/cgi/content/abstract/5/1/68}{{Integrated
  Covariance Estimation using High-frequency Data in the Presence of Noise}},
  Journal of Financial Econometrics 5~(1) (2007) 68--104.
\newblock \href
  {http://arxiv.org/abs/http://jfec.oxfordjournals.org/cgi/reprint/5/1/68.pdf}
  {\path{arXiv:http://jfec.oxfordjournals.org/cgi/reprint/5/1/68.pdf}}, \href
  {http://dx.doi.org/10.1093/jjfinec/nbl011}
  {\path{doi:10.1093/jjfinec/nbl011}}.
\newline\urlprefix\url{http://jfec.oxfordjournals.org/cgi/content/abstract/5/1%
/68}

\bibitem{griffin06}
J.~E. Griffin, R.~C. Oomen, Covariance measurement in the presence of
  non-synchronous trading and market microstructure noise, Journal of
  Econometrics In Press, Corrected Proof.

\bibitem{toth09}
B.~T{\'o}th, J.~Kert{\'e}sz, The epps effect revisited, Quantitative Finance
  9~(7) (2009) 793--802.
\newblock \href {http://dx.doi.org/10.1080/14697680802595668}
  {\path{doi:10.1080/14697680802595668}}.

\bibitem{zhang08}
L.~Zhang, Estimating covariation: Epps effect, microstructure noise, Journal of
  Econometrics In Press.
\newblock \href {http://dx.doi.org/10.1016/j.jeconom.2010.03.012}
  {\path{doi:10.1016/j.jeconom.2010.03.012}}.

\bibitem{barndorff02}
O.~E. Barndorff-Nielsen, N.~Shephard,
  \href{http://ideas.repec.org/p/nuf/econwp/0213.html}{Econometric analysis of
  realised covariation: High frequency covariance, regression and correlation
  in financial economics}, Economics Papers 2002-W13, Economics Group, Nuffield
  College, University of Oxford (2001).
\newline\urlprefix\url{http://ideas.repec.org/p/nuf/econwp/0213.html}

\bibitem{barndorff08}
O.~E. Barndorff-Nielsen, P.~R. Hansen, A.~Lunde, N.~Shephard,
  \href{http://ideas.repec.org/p/nuf/econwp/0810.html}{Multivariate realised
  kernels: consistent positive semi-definite estimators of the covariation of
  equity prices with noise and non-synchronous trading}, Economics Papers
  2008-W10, Economics Group, Nuffield College, University of Oxford (2008).
\newline\urlprefix\url{http://ideas.repec.org/p/nuf/econwp/0810.html}

\bibitem{reno03}
R.~Ren{\`o}, A closer look at the epps-effect, International Journal of
  Theoretical and Applied Finance 6~(1) (2003) 87--102.

\bibitem{muennix09b}
M.~C. M{\"u}nnix, R.~Sch{\"a}fer, T.~Guhr, Compensating asynchrony effects in
  the calculation of financial correlations, Physica A 389~(4) (2010) 767--779.
\newblock \href {http://dx.doi.org/10.1016/j.physa.2009.10.033}
  {\path{doi:10.1016/j.physa.2009.10.033}}.

\bibitem{martens01}
M.~Martens, S.-H. Poon,
  \href{http://www.sciencedirect.com/science/article/B6VCY-43XFHT0-1/2/d81bc0d%
098bd245deef0e9d31f352682}{Returns synchronization and daily correlation
  dynamics between international stock markets}, Journal of Banking \& Finance
  25~(10) (2001) 1805 -- 1827.
\newblock \href {http://dx.doi.org/10.1016/S0378-4266(00)00159-X}
  {\path{doi:10.1016/S0378-4266(00)00159-X}}.
\newline\urlprefix\url{http://www.sciencedirect.com/science/article/B6VCY-43XF%
HT0-1/2/d81bc0d098bd245deef0e9d31f352682}

\bibitem{muennix10a}
M.~C. M{\"u}nnix, R.~Sch{\"a}fer, T.~Guhr, Impact of the tick-size on financial
  returns and correlations, Physica A 389~(21) (2010) 4828--4843,
  http://dx.doi.org/10.1016/j.physa.2010.06.037.
\newblock \href {http://dx.doi.org/10.1016/j.physa.2010.06.037}
  {\path{doi:10.1016/j.physa.2010.06.037}}.

\bibitem{declerck02}
D.~Bourghelle, F.~Declerck, Why markets should not necessarily reduce the tick
  size, Journal of Banking \& Finance 28~(2) (2004) 373 -- 398.
\newblock \href {http://dx.doi.org/10.1016/S0378-4266(03)00136-5}
  {\path{doi:10.1016/S0378-4266(03)00136-5}}.

\bibitem{huang01}
R.~D. Huang, H.~R. Stoll, Tick size, bid-ask spreads, and market structure,
  Journal of Financial and Quantitative Analysis 36~(04) (2001) 503--522.
\newblock \href {http://dx.doi.org/10.2307/2676222}
  {\path{doi:10.2307/2676222}}.

\bibitem{harris91}
L.~Harris, Stock price clustering and discreteness, Review of Financial Studies
  4~(3) (1991) 389--415.
\newblock \href {http://dx.doi.org/10.1093/rfs/4.3.389}
  {\path{doi:10.1093/rfs/4.3.389}}.

\bibitem{kozicki09}
S.~Kozicki, B.~Hoffman,
  \href{http://ideas.repec.org/a/mcb/jmoncb/v36y2004i3p319-38.html}{Rounding
  error: A distorting influence on index data}, Journal of Money, Credit and
  Banking 36~(3) (2004) 319--38.
\newline\urlprefix\url{http://ideas.repec.org/a/mcb/jmoncb/v36y2004i3p319-38.h%
tml}

\bibitem{angel97}
J.~J. Angel,
  \href{http://ideas.repec.org/a/bla/jfinan/v52y1997i2p655-81.html}{Tick size,
  share prices, and stock splits}, Journal of Finance 52~(2) (1997) 655--681.
\newline\urlprefix\url{http://ideas.repec.org/a/bla/jfinan/v52y1997i2p655-81.h%
tml}

\bibitem{onnela09}
J.-P. Onnela, J.~T{\"o}yli, K.~Kaski,
  \href{http://www.sciencedirect.com/science/article/B6TVG-4TPPF19-1/2/282ce63%
e7c76e54d433dc58fb164079f}{Tick size and stock returns}, Physica A: Statistical
  Mechanics and its Applications 388~(4) (2009) 441 -- 454.
\newblock \href {http://dx.doi.org/10.1016/j.physa.2008.10.014}
  {\path{doi:10.1016/j.physa.2008.10.014}}.
\newline\urlprefix\url{http://www.sciencedirect.com/science/article/B6TVG-4TPP%
F19-1/2/282ce63e7c76e54d433dc58fb164079f}

\bibitem{hansen06}
P.~R. Hansen, A.~Lunde,
  \href{http://ideas.repec.org/a/bes/jnlbes/v24y2006p127-161.html}{Realized
  variance and market microstructure noise}, Journal of Business \& Economic
  Statistics 24 (2006) 127--161.
\newline\urlprefix\url{http://ideas.repec.org/a/bes/jnlbes/v24y2006p127-161.ht%
ml}

\bibitem{szpiro98}
G.~G. Szpiro,
  \href{http://www.sciencedirect.com/science/article/B6VCY-41XMJV1-4/2/3bfb801%
7f7acde3b7cebc5d474314946}{Tick size, the compass rose and market
  nanostructure}, Journal of Banking \& Finance 22~(12) (1998) 1559 -- 1569.
\newblock \href {http://dx.doi.org/DOI: 10.1016/S0378-4266(98)00073-9}
  {\path{doi:DOI: 10.1016/S0378-4266(98)00073-9}}.
\newline\urlprefix\url{http://www.sciencedirect.com/science/article/B6VCY-41XM%
JV1-4/2/3bfb8017f7acde3b7cebc5d474314946}

\bibitem{bollerslev86}
T.~Bollerslev,
  \href{http://www.sciencedirect.com/science/article/B6VC0-46VV78N-4/2/ce371da%
ca28a0d38824736988b1d0ef1}{Generalized autoregressive conditional
  heteroskedasticity}, Journal of Econometrics 31~(3) (1986) 307 -- 327.
\newblock \href {http://dx.doi.org/10.1016/0304-4076(86)90063-1}
  {\path{doi:10.1016/0304-4076(86)90063-1}}.
\newline\urlprefix\url{http://www.sciencedirect.com/science/article/B6VC0-46VV%
78N-4/2/ce371daca28a0d38824736988b1d0ef1}

\bibitem{b_cont}
R.~Cont, P.~Tankov, {Financial Modelling with Jump Processes}, second edition
  Edition, CRC Press, Taylor and Francis, 2010, Ch.~7.

\end{thebibliography}

\end{document}